\documentclass{ws-procs975x65}

\def\beq{\begin{equation}}
\def\eeq{\end{equation}}

\begin{document}

\title{Hamilton Geometry\\ - \\Phase Space Geometry from Modified Dispersion Relations}

\author{Christian Pfeifer}

\address{Institute for Theoretical Physics, University of Hannover,\\
	Hannover, 30167, Germany\\
	E-mail: christian.pfeifer@itp.uni-hannover.de}

\author{Leonardo Barcaroli}

\address{Dipartimento di Fisica, Universit`a ”La Sapienza” and
Sez. Roma1 INFN,\\
	Roma, 00185, Italy\\
	E-mail: leonardo.barcaroli@roma1.infn.it}

\author{Lukas K. Brunkhorst}

\address{Center of Applied Space Technology and Microgravity (ZARM), University of Bremen,\\
	Bremen, 28359, Germany\\
	E-mail: lukas.brunkhorst@zarm.uni-bremen.de}

\author{Giulia Gubitosi}

\address{Theoretical Physics Blackett Laboratory, Imperial College,\\
	London, SW7 2AZ, United Kingdom\\
	E-mail: g.gubitosi@imperial.ac.uk}

\author{Niccol\'o Loret}
\address{Institut Ruđer Bošković,\\
	10000 Zagreb, Bijenička cesta 54, Croatia\\
	E-mail: Niccolo.Loret@irb.hr}

\begin{abstract}
Quantum gravity phenomenology suggests an effective modification of the general relativistic dispersion relation of freely falling point particles caused by an underlying theory of quantum gravity. Here we analyse the consequences of modifications of the general relativistic dispersion on the geometry of spacetime in the language of Hamilton geometry. The dispersion relation is interpreted as the Hamiltonian which determines the motion of point particles. It is a function on the cotangent bundle of spacetime, i.e. on phase space, and determines the geometry of phase space completely, in a similar way as the metric determines the geometry of spacetime in general relativity. After a review of the general Hamilton geometry of phase space we discuss two examples. The phase space geometry of the metric Hamiltonian $H_g(x,p)=g^{ab}(x)p_ap_b$ and the phase space geometry of the first order q-de Sitter dispersion relation of the form $H_{qDS}(x,p)=g^{ab}(x)p_ap_b + \ell G^{abc}(x)p_ap_bp_c$ which is suggested from quantum gravity phenomenology. We will see that for the metric Hamiltonian $H_g$ the geometry of phase space is equivalent to the standard metric spacetime geometry from general relativity. For the q-de Sitter Hamiltonian $H_{qDS}$ the Hamilton equations of motion for point particles do not become autoparallels but contain a force term, the momentum space part of phase space is curved and the curvature of spacetime becomes momentum dependent.
\end{abstract}


\bodymatter


\section{Covariant Dispersion Relations, Quantum Gravity Phenomenology and the Geometry of Spacetime} 
One way how quantum gravity is expected to manifest itself in observations is a modification of the standard dispersion relation in metric spacetime geometry \cite{AmelinoCamelia:2008qg,Magueijo:2002xx}. For each observer on a timelike worldline $\gamma$ in spacetime the standard dispersion relation of a freely falling particle is
\begin{equation}\label{eq:1}
E^2=\vec{p}\ {}^2+m^2\,.
\end{equation}
Here $E$ is the energy and  $\vec p$ the spatial momentum the observer associates to the particle, while $m$ is the invariant mass parameter of the particle. With help of the spacetime metric~$g$, its inverse~$g^{-1}$ and the four-momentum of the particle $p$, possibly derived from a Lagrangian $L(x,\dot x)$ that determines the motion of the particle,
$p_a=\frac{\partial}{\partial \dot x^a}L\,,$
the dispersion relation can be written covariantly in terms of the Hamiltonian
\begin{equation}\label{eq:3}
H(x,p)=g^{ab}(x)p_ap_b=m^2\,.
\end{equation}
The relation between equation (\ref{eq:1}) and equation (\ref{eq:3}) is given by the expansion of the latter in an orthonormal frame of the metric associated to the observer on the worldline $\gamma$. This short analysis of the dispersion relation of freely falling particles shows clearly that the dispersion relation is closely intertwined with the geometry of spacetime, i.e. with the spacetime metric~$g$. Thus any change of the dispersion relation of freely falling particles induces a modification of the geometry of spacetime and, the other way around, changes in the geometry of spacetime alter the covariant dispersion relation.

Several models of quantum gravity phenomenology suggest that quantum gravity effects, influencing the motion of high energetic particles, manifest themselves in a modification of the dispersion relation~(\ref{eq:1})
\begin{equation}\label{eq:4}
E^2=\vec{p}\ {}^2+m^2+\ell f(E,\vec{p})+\mathcal{O}(\ell^2),
\end{equation}
where $\ell$ is the Planck length, the scale at which quantum gravity effects should become relevant \cite{Amelino-Camelia:2014rga,Girelli:2006fw}. Immediately a question arises: How can one cast such modified dispersion relations in a covariant form and what is the geometric structure of the underlying spacetime manifold? As soon as non-quadratic terms in the energy and the spatial momenta appear the geometry can not be metric spacetime geometry. What is however always possible is to interpret the modified dispersion relation as level sets of a Hamiltonian on phase space
\begin{equation}\label{eq:5}
H(x,p)=m^2\,,
\end{equation}
To analyse the geometry that is determined by a general Hamiltonian on phase space we employ Hamilton geometry. The latter enables us to compare the geometries on phase space induced by different Hamiltonians, for example by the metric Hamiltonian, displayed in equation (\ref{eq:3}) and by the q-de Sitter Hamiltonian suggested from quantum gravity phenomenology \cite{Marciano:2010gq}
\begin{equation}
H_{qDS}(x,p)=p_0^2-p_1^2(1-2 h x^0)-\ell p_0 p_1^2(1-2hx^0)+\mathcal{O}(\ell^2)\,.
\end{equation}

\section{Hamilton Geometry}
The geometry of Hamilton spaces, mathematically worked out in \cite{Miron} and adopted for the application to quantum gravity phenomenology by the authors \cite{Barcaroli:2015xda}, tells us exactly how we can derive the geometry of phase space from a Hamilton function and its derivatives, once the symplectic structure on phase space is fixed.	 Most importantly Hamilton geometry gives a precise formalism to obtain the curved geometry of spacetime and momentum space as parts of phase space. The fundamental idea is to introduce a connection on phase space which splits the directions on phase space into directions along momentum space and directions which can be interpreted as directions along configuration space, i.e. spacetime. This split then allows to define connections which determine the geometry of momentum space and spacetime each seen as part of phase space. Remarkably the split of phase space into spacetime and momentum space as well as the construction of the geometries is invariant  with respect to spacetime diffeomorphisms. This means the usual spacetime diffeomorphism invariance as known from metric spacetime geometry is realised in the Hamilton geometry framework. Comparing the general construction for an arbitrary Hamiltonian $H$ and a metric Hamiltonian $H_g=g^{ab}(x)p_ap_b$ we find

\begin{center}
	\begin{tabular}{ | c | c | c | }
		\hline
		& Hamilton geometry & Metric geometry \\ \hline 
		Fundamental & $H(x,p)$ a function on & $g(x)$ a metric on  \\ 
		object & phase space & spacetime  \\ \hline
		Phase space connection  & $N_{ab}(x,p)$ & $-\Gamma^a{}_{bc}(x)p_a$ \\ \hline
		Spacetime connection & $\Gamma^{\delta a}{}_{bc}(x,p)$ & $\Gamma^a{}_{bc}(x)$  \\ \hline
		Momentum space connection & $C^{ab}{}_c(x,p)$ & $0$  \\ \hline
	\end{tabular}
\end{center}
The unique symmetric Hamilton metric compatible connection coefficients are
\begin{eqnarray}
\Gamma^a{}_{bc}(x)&=&\frac{1}{2}g^{aq}(\partial_bg_{qc}+\partial_ag_{qb}-\partial_qg_{bc})\\
N_{ab}(x,p)&=&\frac{1}{4}\big(\{g^H_{ab}, H\}+g^H_{ai}\partial_b\bar{\partial}^i H+g^H_{bi}\partial_a\bar{\partial}^i H\big),\\
\Gamma^{\delta a}{}_{bc}(x)&=&\frac{1}{2}g^{aq}(\delta_bg_{qc}+\delta_ag_{qb}-\delta_qg_{bc}),\quad C^{ab}{}_c(x,p)=-g^H_{qc}\bar{\partial}^qg^{Hab}\,,
\end{eqnarray}
where we used the abbreviations $\frac{\partial}{\partial x^a}=\partial_a$, $\delta_a=\partial_a-N_{ab}\bar{\partial}^b$, the Poisson bracket $\{\cdot,\cdot\}$ between two phase space functions and the Hamilton metric
\begin{equation}
g^{Hab}=\frac{1}{2}\frac{\partial}{\partial p_a} \frac{\partial}{\partial p_b}H=\frac{1}{2}\bar{\partial}^a\bar{\partial}^b H\,.
\end{equation}
It is remarkable that for the metric Hamiltonian the momentum space connection is automatically flat and the spacetime connection is given by the Christoffel-Symbols of the Levi-Civita connection. For a generic Hamiltonian, for which the third derivative with respect to the momenta does not vanish, the momentum space connection is non-vanishing and yields a non-trivial curvature of momentum space, as we will demonstrate explicitly at the example of the q-de Sitter dispersion relation below. Thus the metric geometry of phase space employed in general relativity is a very distinguished one compared to geometries derived from general dispersion relations. When we now study the Hamilton equations of motion of point particles in Hamilton geometry we find that they can be understood as the autoparallel equations of the phase space connection with source term and the standard relation between momenta and velocities
\begin{equation}\label{eq:hamauto}
\dot p_a + \partial_a H =0,\quad \dot x^a=\bar{\partial}^aH\quad \Leftrightarrow\quad \dot p_a + N_{ab}\bar\partial^bH=-\delta_aH,\quad \dot x^a=\bar{\partial}^aH\,.
\end{equation}
The part $\dot p_a + N_{ab}\bar\partial^bH$ of the $\dot p$ equations of motion are the autoparallel equations of the phase space connection and the $-\delta_a H$ is a force like term which drags particles away from autoparallel motion. An important result in this investigation is that for homogeneous Hamiltonians $H(x,\lambda p)=\lambda ^r H(x,p)$ the source term vanishes and point particles, propagating according to Hamilton's equations of motion, move on autoparallels of the Hamilton geometry. In particular for $H=g^{ab}(x)p_ap_b$ the Hamilton equations of motion become the geodesic equation of the metric $g$ and the usual duality relation between velocities and momenta.

\section{Hamilton Geometry of q-de Sitter Phase Space} 
With help of the Hamilton geometry of phase space we now analyse the phase space geometry of the q-de Sitter dispersion relation from quantum gravity phenomenology. It falls in the class of first order cubic modifications of the metric dispersion relation of general relativity
\begin{equation}
H_{qDS}(x,p)=g^{ab}(x)p_ap_b+\ell G^{abc}p_ap_bp_c=p_0^2-p_1^2(1-2 h x^0)-\ell p_0 p_1^2(1-2hx^0)\,.
\end{equation}
It includes the Planck length $\ell$ and the expansion rate $h$ as parameters. For $h=0$ the q-de Sitter dispersion relation becomes the $\kappa$-Poincar\'e dispersion relation, which is often and long discussed in the literature \cite{Majid:1994cy}, while for $\ell=0$ we obtain the standard $1+1$ dimensional de Sitter spacetime dispersion relation. In this first order in $\ell$ expansion the geometry we obtain can be interpreted as multi-tensor geometry since it is derived from the metric $g^{ab}$ and the tensor $G^{abc}$. The Hamilton metric and its inverse can easily be obtained
\begin{eqnarray}
g^{Hab}=\frac{1}{2}\bar{\partial}^a\bar{\partial}^bH_{qdS} &=&g^{ab}+\ell 3 G^{abc}p_c,\\
&=&\left(\begin{array}{lll}1 &\;\;& -\ell p_1 (1+2 h x^0) \\
-\ell p_1 (1+2 h x^0) &\;\;&- (1+2 h x^0)(1+ \ell p_0) \\ \end{array}\right)\,,\\
g^{H}{}_{ab}& =&g_{ab}-\ell 3 g_{aj}g_{bi} G^{ijc}p_c\\
&=&\left(\begin{array}{lll}1 &\;\;& -\ell p_1 \\ -\ell p_1 &\;\;& -(1-2 h x^0)(1- \ell p_0)\end{array}\right)\,.
\end{eqnarray}
The phase space connection coefficients can be directly calculated from their defining equation
\begin{eqnarray}
N_{ab}&=&-p_q\Gamma^q{}_{ab}+\ell \frac{3}{4}p_cp_d\big(g_{qb}\nabla_a G^{qcd}+g_{qa}\nabla_b G^{qcd}-2 g_{ma}g_{nb}g^{qc}\nabla_qG^{dmn}\big)\\
&=&\left(
\begin{array}{cc}
h  \ell p_1^2& h p_1 \\
h p_1 & h p_0 (1-\ell p_{0})\\
\end{array}
\right)\,.
\end{eqnarray}
From this expression it becomes very clear that there is a contribution to the connection coefficients from the metric part of the $H_{qDS}$ and a contribution from the third order polynomial part which comes with the parameter $\ell$. The $\dot p$ Hamilton equations of motion in this example can be written as
\begin{eqnarray}
\dot p_0+ 2 h p_1^2+\mathcal{O}(\ell^2)=-2 h \ell p_0 p_1^2+\mathcal{O}(\ell^2),\quad \dot p_1+\ell h p_1^3 +\mathcal{O}(\ell^2)=\ell h p_1^3 +\mathcal{O}(\ell^2)\,.\label{eq:hamautoex2}
\end{eqnarray}
In the geometry defined by the Hamiltonian $H_{qDS}$ the autoparallels of the connection are curves for which satisfy the right hand side of equations (\ref{eq:hamautoex2}) vanishes. Thus the Hamilton equations of motion are not autoparallels in the geometry but include force-like terms which drag the particles away from autoparallel motion. The cause of this-force like terms lies in the different homogeneities of the different terms in the Hamiltonian with respect to the momenta. Finally the geometry of spacetime and momentum space as parts of phase space is determined by the connection coefficients $\Gamma^{\delta a}{}_{bc}$  and $C^{ab}{}_c$ defined in the previous section. The non-vanishing coefficients are
\begin{eqnarray}
\Gamma^{\delta0}{}_{01}&=&-  h \ell p_{1},\quad \Gamma^{\delta0}{}_{11}=-h(1-2\ell p_{0}) \\
\Gamma^{\delta1}{}_{00}&=&- h \ell p_1,\quad \Gamma^{\delta1}{}_{01}=-h ,\quad\Gamma^{\delta1}{}_{11}=\frac{3}{2}  h \ell p_{1}\,,
\end{eqnarray}
and
\begin{eqnarray}\label{eq:Cel1}
C^{01}{}_1=-\frac{\ell}{2}(1-2 h x^0)+\mathcal{O}(\ell^2),\quad C^{11}{}_0=\frac{\ell}{2}(1-2 h x^0)+\mathcal{O}(\ell^2)\,.\label{eq:Cel2}
\end{eqnarray} 
The curvature of spacetime $R^a{}_{bcd}$, measuring the non-triviality of parallel transport along spacetime, takes the form
\begin{eqnarray}
R^{Ha}{}_{bcd}&=&\delta_c\Gamma^{\delta a}{}_{bd}-\delta_d\Gamma^{\delta a}{}_{bc} +\Gamma^{\delta a}{}_{ci}\Gamma^{\delta i}{}_{bd}-\Gamma^{\delta a}{}_{di}\Gamma^{\delta i}{}_{bc}-R_{icd}C^{qi}{}_{b}\\
&=&R^a{}_{bcd}(x)+\ell p_q(\nabla_c\gamma^{qa}{}_{bd}-\nabla_d\gamma^{qa}{}_{bc}+\frac{3}{2}R^q{}_{rcd}G^{ra}{}_b)
\end{eqnarray}
where $R_{icd}=[\delta_c,\delta_d]_i$, $\gamma^{qa}{}_{bd}$ are the first order corrections in $\ell$ to the Christoffel symbols derived from the tensor components $G^{abc}$ and $R^a{}_{bcd}$ is the Riemann tensor of the metric $g$. This expression of the spacetime curvature contains parts proportional to $h$ and $\ell$ and and nicely demonstrates the covariance of the framework as well as that there appears a momentum dependence in the spacetime curvature due to the non-quadratic dependence of the dispersion relation on the momenta. The curvature of phase space for this Hamiltonian is of order $\ell^2$
\begin{equation}
Q^{cab}{}_{q}=\bar\partial^a C^{bc}{}_{q}-\bar\partial^b C^{ac}{}_{q}+C^{am}{}_{q}C^{bc}{}_{m}-C^{bm}{}_{q}C^{ac}{}_{m}=\mathcal{O}(\ell^2)\,.
\end{equation}
The expression for the curvature of spacetime and momentum space of the q-de Sitter dispersion relation demonstrate nicely the strengths of the Hamilton geometry formalism. In the metric dispersion relation limit $\ell=0$ the momentum space connection coefficients $C$ vanish as well as the momentum space curvature $Q$ while the spacetime curvature is non-vanishing and given by the usual Riemann curvature tensor of metric geometry. In the $\kappa$-Poincare limit $h=0$  the opposite happens, the spacetime curvature vanishes but the momentum space is naturally curved. In the general case curvature is present in both: position and momentum space.

\section{Conclusion and Outlook} 
Hamilton geometry enables us to derive the geometry of phase space from dispersion relations and to compare different phase space geometries. In particular we can  identify the geometry of spacetime and the geometry of momentum space as parts of phase space naturally in a spacetime diffeomorphism invariant way. For generic Hamiltonians it turns out that both are curved. Hamilton geometry is an optimal framework to study dispersion relation induced phase space geometries with curved position and momentum space, with flat position and curved momentum space as well as with curved position and a flat momentum space.

The next important step in the application of Hamilton geometry to quantum gravity modified dispersion relation is to understand how the addition of momenta can be realised with help of the geometric tools introduced here.

\end{document}